\documentclass[a4paper,fleqn]{cas-sc}
\usepackage{comment}
\usepackage[numbers]{natbib}
\usepackage{gensymb}
\usepackage[utf8]{inputenc}
\usepackage{ulem}
\usepackage{etoolbox}
\DeclareUnicodeCharacter{22EF}{\dots}

\usepackage{soul}
\usepackage{siunitx}
\newcommand{\svert}{\vert}

\def\tsc#1{\csdef{#1}{\textsc{\lowercase{#1}}\xspace}}
\tsc{WGM}
\tsc{QE}
\tsc{EP}
\tsc{PMS}
\tsc{BEC}
\tsc{DE}

\begin{document}
\let\WriteBookmarks\relax
\def\floatpagepagefraction{1}
\def\textpagefraction{.001}
\shorttitle{Sodium-Decorated P-C$_3$N}
\shortauthors{Laranjeira et~al.}

\title [mode = title]{Sodium-Decorated P-C$_3$N: A Porous 2D Framework for High-Capacity and Reversible Hydrogen Storage}

\author[1]{José A. S. Laranjeira}
\cormark[1]
\cortext[cor1]{Corresponding author}
\affiliation[1]{
organization={Modeling and Molecular Simulation Group},
addressline={São Paulo State University (UNESP), School of Sciences}, 
city={Bauru},
postcode={17033-360}, 
state={SP},
country={Brazil}}
\credit{Conceptualization of this study, Methodology, Review and editing, Investigation, Formal analysis, Writing -- review \& editing, Writing -- original draft}

\author[1]{Nicolas F. Martins}
\credit{Conceptualization of this study, Methodology, Review and editing, Investigation, Formal analysis, Writing -- review \& editing, Writing -- original draft}

\author[2]{Kleuton A. L. Lima}
\affiliation[2]{
organization={Department of Applied Physics and Center for Computational Engineering and Sciences},
addressline={State University of Campinas}, 
city={Campinas},
postcode={13083-859}, 
state={SP},
country={Brazil}}
\credit{Conceptualization of this study, Methodology, Review and editing, Investigation, 
Formal analysis, Writing -- review \& editing, Writing -- original draft}

\author[3]{Lingtao Xiao}
\affiliation[3]{
organization={School of Materials Science and Engineering},
addressline={Chongqing University of Arts and Sciences}, 
city={Chongqing},
postcode={402160}, 
country={China}}
\credit{Conceptualization of this study, Methodology, Review and editing, Investigation, 
Formal analysis, Writing -- review \& editing, Writing -- original draft}

\author[3]{Xihao Chen}
\credit{Conceptualization of this study, Methodology, Review and editing, Investigation, 
Formal analysis, Writing -- review \& editing, Writing -- original draft}

\author[4,5]{Luiz A. Ribeiro Junior}
\affiliation[4]{
organization={Institute of Physics},
addressline={University of Brasília}, 
city={Brasília },
postcode={70910‑900}, 
state={DF},
country={Brazil}}
\affiliation[5]{
organization={Computational Materials Laboratory, LCCMat, Institute of Physics},
addressline={University of Brasília}, 
city={Brasília },
postcode={70910‑900}, 
state={DF},
country={Brazil}}
\credit{Conceptualization of this study, Methodology, Review and editing, Investigation, 
Formal analysis, Writing -- review \& editing, Writing -- original draft}
\author[1]{Julio R. Sambrano}
\credit{Conceptualization of this study, Methodology, Review and editing, Investigation, 
Formal analysis, Writing -- review \& editing, Writing -- original draft}


\begin{abstract}
The development of reversible hydrogen storage materials has become crucial for enabling carbon-neutral energy systems. Based on this, the present work investigates the hydrogen storage on the sodium-decorated P-C$_3$N (Na@P-C$_3$N), a porous carbon nitride monolayer recently proposed as a stable semiconductor. First-principles calculations reveal that Na atoms preferentially adsorb with an adsorption energy of -4.48~eV, effectively suppressing clusterization effects. Upon decoration, the system becomes metallic, while \textit{ab initio} molecular dynamics simulations confirm the thermal stability of Na@P-C$_3$N at 300~K. Hydrogen adsorption on Na@P-C$_3$N occurs through weak physisorption, with energies ranging from -0.18 to -0.28~eV, and desorption temperatures between 231 and 357~K. The system can stably absorb 16 H$_2$ molecules per unit cell, corresponding to a gravimetric storage capacity of 9.88~wt\%, surpassing the U.S. Department of Energy target. These results demonstrate that Na@P-C$_3$N is a promising candidate for lightweight, stable, and reversible hydrogen storage.
\end{abstract}


\begin{highlights}
\item P-C$_3$N is a stable 2D porous semiconductor with high Na adsorption energy.
\item Na decoration induces metallicity while preserving structural integrity.
\item Up to 16 H$_2$ molecules adsorb with reversible energies (-0.18 to -0.28 eV).
\item Hydrogen storage reaches 9.88 wt\%, surpassing DOE 2025 target.
\item AIMD and thermodynamic analysis confirm ambient-condition reversibility.
\end{highlights}


\maketitle

\section{Introduction}

The transition toward a sustainable energy future has intensified the search for clean, safe, and efficient energy carriers~\cite{Chu2012}. Among the various alternatives, hydrogen stands out due to its high gravimetric
energy density and environmental friendliness, making it a promising candidate for next-generation energy storage systems~\cite{Evro2024,Ghotia2025}. However, developing suitable materials with the ability to store
hydrogen under reversible conditions remains one of the most pressing challenges in the field~\cite{Rasul2022,Durbin2013}.

Two-dimensional (2D) materials have emerged as attractive platforms for hydrogen storage due to their high surface area, tunable electronic properties, and mechanical
flexibility~\cite{Fan2021,Qin2023,rahimi2025lithium,zhang2024decorated,praveer2025density,ma2024ti,miao2024ultrahigh,ahmed2025unveiling,ahmed2023ab,ahmed2024dft,ahmed2024first}. Recent advances have identified a wide
variety of metal-decorated 2D materials as up-and-coming candidates for reversible hydrogen storage. For instance, Li-decorated C$_4$N monolayers~\cite{ZHANG202132936} exhibit outstanding structural and thermal
stability, with Li atoms uniformly adsorbed at hollow sites. This system can store up to six H$_2$ molecules per Li atom, reaching a gravimetric hydrogen storage capacity of 8.00~wt\% and moderate adsorption energies
around $-0.281~{\rm eV}$ per H$_2$.

Another example is the Sc$_3$N$_2$ MXene monolayer, where surface functionalization with Li and Na significantly enhances its H$_2$ storage performance. Each Li (or Na) site is capable of adsorbing
multiple H$_2$ molecules, with gravimetric capacities of 5.90~wt\% (Li) and 5.63~wt\% (Na). Hydrogen molecules are physisorbed through a combination of van der Waals interactions~\cite{TAYYAB2025115489}.

Li-decorated hexagonal B$_2$S$_3$ nanosheets have also demonstrated efficient hydrogen uptake, achieving a storage capacity of 6.35~wt\%~\cite{HUZAIFA2025114915}. The Li atoms preferentially adsorb at the
bridge sites between B and S atoms, promoting uniform dispersion and reducing clustering. Hydrogen adsorption occurs through weak chemisorption with average binding energies of $-0.21$~eV/H$_2$, and the system
exhibits hydrogen release around 210~K.

Finally, Ti-decorated orthorhombic B$_2$N$_2$ (o-B$_2$N$_2$) monolayers offer a promising hydrogen storage capacity of 11.21~wt\%~\cite{CHODVADIYA2024958}. Ti atoms are strongly anchored at the
N--B--N bridge sites, where each Ti can adsorb up to five H$_2$ molecules through a combination of electrostatic, polarization, and Kubas-type interactions. The average adsorption energy is around $-0.26$~
eV/H$_2$, which is optimal for reversible storage. 

Inside this framework, Tan \textit{et al.}~\cite{Tan2024} explore the potential of two-dimensional (2D) carbon nitride monolayers via density functional theory (DFT) calculations. They introduce three theoretically stable
monolayers --- T-C$_3$N, P-C$_3$N, and PH-C$_5$N$_4$ --- each featuring tetragonal lattices. These three nanosheets are shown to behave as intrinsic semiconductors. The monolayers also demonstrate
directionally dependent (anisotropic) carrier mobilities and strong optical absorption in both the visible and near-infrared spectral ranges, pointing to their suitability for optoelectronic devices. Importantly, other
porous 2D-CN frameworks, like graphitic carbon nitrides (g-CN)~\cite{lu2025breaking,zhang2023high,wang2022novel,fang2024pursuit,fang2024theoretical,fang2024coordination,fang2024single}, have also been reported with
promising potential for developing novel energy storage applications. 

Here, our focus lies on P-C$_3$N, which exhibits an indirect band gap of 0.96 eV at HSE06 level~\cite{Tan2024}. The material demonstrates promising carrier mobility, exceeding 10$^3$
cm$^2\,$V$^{-1}\,$s$^{-1}$, making it suitable for high-performance electronic applications. Structurally, this monolayer features an arrangement of pyrrolopyrrole-like motifs, forming a porous lattice
composed of \textit{sp}-hybridized nitrogen atoms and carbon octagonal rings. This unique atomic configuration could offer favorable sites for metal atom adsorption, suggesting potential applications in hydrogen storage
via metal decoration. 

Motivated by these insights, we explore the hydrogen storage \,  performance of sodium-decorated P-C$_3$N (Na@P-C$_3$N) through first-principles DFT calculations. Sodium, a light and abundant alkali metal, has
been shown to exhibit favorable interactions with 2D substrates, allowing the formation of stable adsorption sites while facilitating electron donation to H$_2$ molecules~\cite{Chen2025,Martins2024}. Here, we
demonstrate that Na@P-C$_3$N meets and exceeds the U.S. Department of Energy (DOE) 2025 targets, combining strong adsorption, excellent thermal stability, and reversibility under near-ambient conditions.

\section{Methodology}
\label{sec:methods}

\begin{figure*}[!htb]
    \centering
    \includegraphics[width=\linewidth]{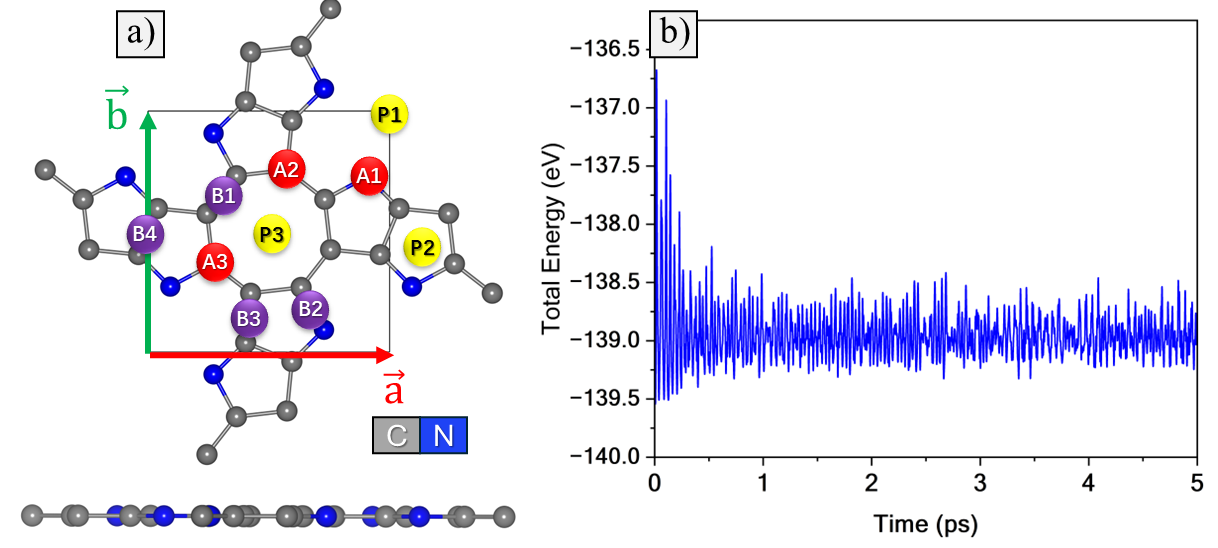}
    \caption{(a) Top and side views of the optimized P-C$_3$N monolayer. Adsorption sites are labeled as atomic (A1--A3), bridge (B1--B4), and pore-centered (P1--P3). Carbon and nitrogen atoms are shown in gray and blue, respectively. (b) The total energy profile from the AIMD simulation at 300~K for 5~ps demonstrates thermal stability. The lattice vectors are $\vec{a}=\vec{b}=7.03$ \r{A}.}
    \label{fig:structure}
\end{figure*}

All calculations were performed within the framework of density functional theory (DFT) as implemented in the Vienna \textit{ab initio} simulation package (VASP)~\cite{Kresse1996,Kresse1999}. The exchange--correlation
interactions were treated using the generalized gradient approximation (GGA) in the Perdew--Burke--Ernzerhof (PBE) formulation~\cite{Perdew1996}. Core--valence interactions were described by the projector augmented-wave
(PAW) method~\cite{Blochl1994}. A plane-wave energy cutoff of 520~eV was adopted for all calculations.

A vacuum spacing of 15~\AA~was applied along the out-of-plane direction to avoid spurious interactions between periodic images. The Brillouin zone was sampled using a $\Gamma$-centered Monkhorst--Pack mesh of
$5~\times~5~\times~1$. Atomic positions were relaxed using the conjugate gradient algorithm until the forces on each atom were less than 0.01~eV/\AA\ and the energy change between steps was below $1 \times 10^{-5}~{\rm eV}$. Dispersion interactions
were included using the DFT-D2 method of Grimme~\cite{Grimme2006}.

All calculations were performed using the GGA/PBE scheme. While hybrid functionals, such as HSE06, provide more accurate bandgap estimations, PBE was selected due to its proven reliability in describing adsorption behavior
and charge distribution in 2D systems, offering a lower computational cost. The bandgap value of 0.96 eV reported for pristine P-C$_3$N at the HSE06 level is mentioned here from previous literature~\cite{Tan2024}
for reference only.

To investigate the thermal stability of the studied systems, \textit{ab initio} molecular dynamics (AIMD) simulations were carried out within the NVT ensemble using the Nos\'e thermostat~\cite{Hoover1985}, at a temperature
of 300~K. Each simulation ran for 5~ps with a time step of 0.5~fs to assess the thermal stability of both pristine and sodium-decorated P-C$_3$N systems.

To evaluate the interaction of hydrogen molecules with Na@P-C$_3$N, H$_2$ molecules were incrementally adsorbed near each sodium atom. The adsorption energy per H$_2$ was computed using:
\begin{equation}
E_{\text{ads}} = \frac{1}{n} \left[ E_{\text{Na@P-C}_3\text{N}+n\text{H}_2} - E_{\text{Na@P-C}_3\text{N}} - nE_{\text{H}_2} \right],
\end{equation}
 where $E_{\text{Na@P-C}_3\text{N}+n\text{H}_2}$ is the total energy of the system with $n$ adsorbed H$_2$ molecules, $E_{\text{Na@P-C}_3\text{N}}$ is the energy of the clean Na-decorated surface, and $E_{\text{H}_2}$ is the energy of an isolated hydrogen molecule.

The desorption temperature ($T_{\text{des}}$) was estimated using the van't Hoff relation~\cite{Durbin2013,chodvadiya2024exploring}:
\begin{equation}
T_{\text{des}} = \frac{\lvert E_{\text{ads}}\rvert }{k_B\Delta S/R},
\end{equation}
 where $k_{\mathrm{B}}$ is the Boltzmann constant $(1.380 \times 10^{-23}~\mathrm{J}\,\mathrm{K}^{-1})$, $\Delta S$ is the change in entropy of hydrogen from gas to liquid phase $(75.44~\mathrm{J}\,\mathrm{mol}^{-1}\,\mathrm{K}^{-1})$, and $R$ is the universal gas constant $(8.314~\mathrm{J}\,\mathrm{mol}^{-1}\,\mathrm{K}^{-1})$.

The hydrogen storage capacity (HSC) in weight percentage was calculated as:
\begin{equation}
\text{HSC} (\text{wt\%}) = \frac{ n_{\text{H}} M_{\text{H}}}{n_{\text{H}} M_{\text{H}} + n_{\text{C}} M_{\text{C}} + n_{\text{N}} M_{\text{N}} + n_{\text{Na}} M_{\text{Na}}},
\end{equation}
where $n_X$ and $M_X$ represent the number of atoms and molar masses of element $X$ ($X$ = \text{H}, \text{C}, \text{N}, and \text{Na}), respectively.

Charge transfer analyses were performed using the Bader scheme \cite{Tang2009}, and charge density difference (CDD) plots were generated to visualize the redistribution of electrons. The CDD was computed using:
\begin{equation}
\Delta \rho = \rho_{\text{complex}} - \rho_{\text{substrate}} - \rho_{\text{adsorbate}},
\end{equation}
 where $\rho_{\text{complex}}$ is the charge density of the combined system, and the other terms represent the densities of the isolated components in their relaxed configurations.

Thermodynamic viability under varying pressure and temperature conditions was evaluated using the grand canonical partition function:
\begin{equation}
Z = 1 + \sum_{i=1}^{n} \exp \left( -\frac{E_{\text{ads}}^i - \mu}{k_{\text{B}} T} \right),
\end{equation}
 where $\mu$ is the chemical potential of hydrogen gas, $k_{\text{B}}$ is the Boltzmann constant, and $T$ is the temperature.

\section{Results and discussion}

\subsection{Structural and Electronic Properties of P-C$_3$N}

\begin{figure*}[!htb]
    \centering
    \includegraphics[width=0.7\linewidth]{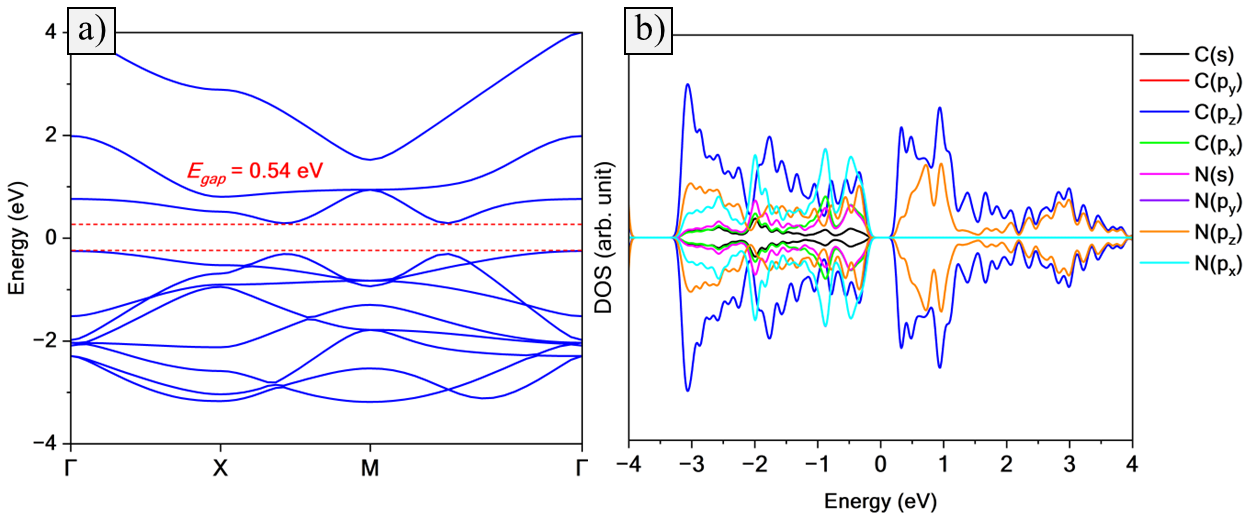}
    \caption{(a) Electronic band structure of P-C$_3$N computed at the PBE level, showing a direct band gap of 0.54~eV at the $\Gamma$ point. (b) PDOS decomposed by orbital contributions of C and N atoms, emphasizing $p_z$-dominated states near the Fermi level.}
    \label{fig:band_dos}
\end{figure*}

Figure~\ref{fig:structure}a shows the top and side views of the optimized P-C$_3$N monolayer, along with the labeling of high-symmetry adsorption sites for subsequent sodium decoration. The structure exhibits a periodic 2D porous framework composed of pyrrolopyrrole motifs arranged in a square lattice (space group P4/m, No. 83). The lattice vectors are $\vec{a}=\vec{b}=7.03$ \r{A}. Four non-equivalent bond lengths were identified: $l_1$ = 1.40 (C--N), $l_2$ = 1.32 (C--N), $l_3$ = 1.49 (C--C), and $l_4$ = 1.40 (C--C). Three classes of adsorption sites are identified: atomic sites (A1--A3, red) positioned above individual atoms, bridge sites (B1--B4, purple) between two neighboring atoms, and pore-centered sites (P1--P3, yellow).

AIMD simulations were conducted at 300~K for 5~ps to assess the thermal robustness. The total energy evolution of the pristine P-C$_3$N system is shown in~\ref{fig:structure}b. The system rapidly stabilizes after a brief initial fluctuation, with energy oscillations confined to a narrow range around $-139~{\rm eV}$. No reconstruction or phase transitions are observed throughout the simulation, confirming the thermal stability of the P-C$_3$N monolayer near ambient conditions.

The electronic structure of the pristine P-C$_3$N was examined by band structure and projected density of states (PDOS), as represented in Figure~\ref{fig:band_dos}. The band structure (Figure~\ref{fig:band_dos}a) reveals an indirect band gap of 0.54~eV at the DFT/PBE level, which is consistent with the value reported by Tan \textit{et al.}~\cite{Tan2024} (0.51 eV). It is noteworthy that along the $ X \rightarrow M $ and $ M \rightarrow \Gamma $
directions, the electronic bands exhibit a nearly quadratic dispersion, indicative of high carrier mobility. This behavior aligns with previous findings~\cite{Tan2024}, which demonstrated that P-C$_3$N significantly
outperforms T-C$_3$N, PH-C$_5$N$_4$, and g-C$_3$N$_4$ in terms of carrier transport properties.

The PDOS shown in Figure~\ref{fig:band_dos}b provides an orbital-resolved analysis of the electronic structure. The valence band is primarily dominated by C($p_z$) states, with notable contributions from N($s$) and N($p$) orbitals, while C($s$) and C($p$) (excluding $p_z$) contribute only marginally. Near the valence band maximum (VBM), N($p_z$) orbitals exhibit the highest contribution. In the conduction band region close to the band edge, only N($p_z$) and C($p_z$) orbitals contribute, highlighting the presence of delocalized $\pi$-type states, which are responsible for the higher band dispersion in this energy range. The symmetrical distribution of PDOS contributions across spin channels confirms the non-magnetic nature of the pristine material.

\subsection{Sodium Decoration on P-C$_3$N}

To evaluate the ability of P-C$_3$N to host alkali metals, we investigated the adsorption of a single Na atom at various sites, such as atomic (A1--A3), bridge (B1--B4), and pore-centered positions (P1--P3), as illustrated previously in Figure~\ref{fig:structure}a. The adsorption energy (E$_{ads}$) was calculated for each configuration. The structural relaxation revealed a clear tendency for the Na atom to migrate toward the center of the pore (P1), regardless of the initial site. All configurations converge to the same final geometry with an identical adsorption energy of $-4.48~{\rm eV}$. Based on this analysis, the P1 site was selected as the primary location for Na decoration, with Na atoms adsorbed both above and below the plane. Each Na atom is four-fold coordinated by surrounding $N$ atoms. Owing to the unique porous framework of P-C$_3$N, additional adsorption sites are available --- specifically, the octagonal carbon rings (P3 sites) --- which provide further capacity for Na decoration. Regardless of the occupancy of the P1 site, Na atoms can also be adsorbed stably at the P3 sites, resulting in a maximum adsorption capacity of four Na adatoms per P-C$_3$N unit cell.

To further assess the anchoring strength of sodium atoms, we estimated the sodium binding energy (E$_b$) using the standard formulation:
\begin{equation*}
E_b = \frac{E_{\mathrm{Na}@\mathrm{P-C3N}} - E_{\mathrm{P-C3N}} - m E_{\mathrm{Na}}^{\mathrm{bulk}}}{m},
\end{equation*}
 where $E_{\mathrm{Na}@\mathrm{P-C3N}}$ and $E_{\mathrm{P-C3N}}$ are the total energies of the decorated and pristine systems, respectively, $E_{\mathrm{Na}}^{\mathrm{bulk}}$ is the cohesive energy of bulk Na (1.13 eV), and $m = 4$ is the number of Na atoms
per unit cell. The estimated binding energy is approximately 3.82 eV per Na atom. This strong binding confirms the stability of Na decoration and rules out spontaneous clustering. This approach is consistent with previous reports~\cite{ma2025and,zhang2025density}.
 
\begin{figure}[!htb]
    \centering
    \includegraphics[width=0.3\linewidth]{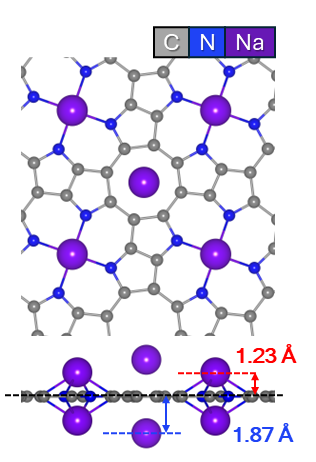}
    \caption{Top and side views of the Na-decorated P-C$_3$N monolayer. Sodium atoms (purple) are located above the pore center with a height of 1.23~\r{A} from the topmost atomic layer and 1.87~\r{A} from the underlying N atoms.}
    \label{fig:na_adsorption}
\end{figure}

Figure~\ref{fig:na_adsorption} illustrates the optimized Na@P-C$_3$N configuration from the top and side views. Na atoms are stabilized above the center of the pore, located at a vertical distance of 1.23~\r{A} from the monolayer plane at the P1 site and 1.87~\r{A} for the P3 site. The difference in the distances between the Na adatoms and the monolayer is associated with the variation in adsorption strength at the respective sites. The Na atoms adsorbed at the P1 site exhibit stronger interactions with the monolayer compared to those adsorbed at the P3 site. This final arrangement minimizes repulsion and maximizes electrostatic stabilization of the electronegative nitrogen. The absence of distortion in the underlying P-C$_3$N lattice confirms that Na adsorption is non-disruptive and structurally favorable.

\begin{figure*}[!htb]
    \centering
    \includegraphics[width=0.7\linewidth]{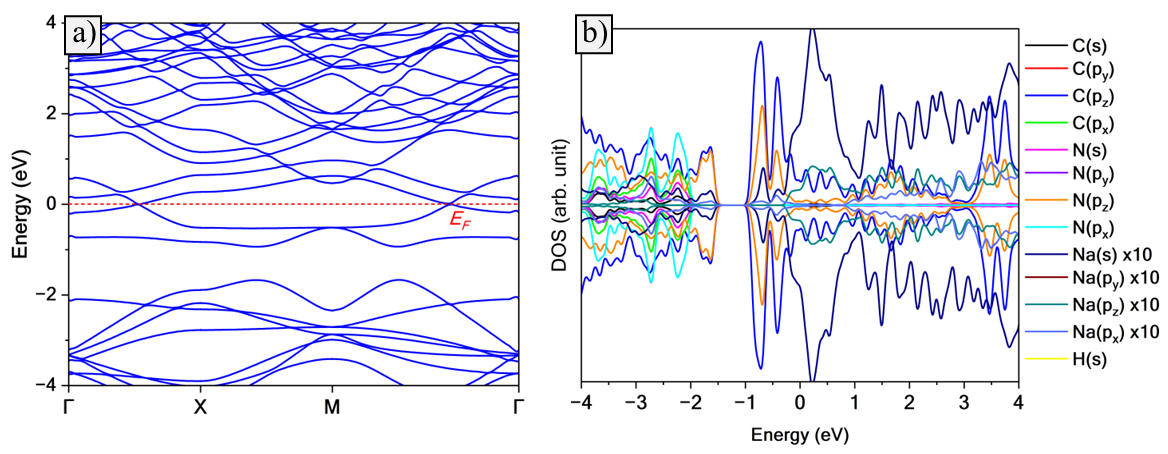}
    \caption{(a) Band structure and (b) projected density of states (PDOS) of Na@P-C$_3$N. Sodium donation shifts the Fermi level into the conduction band, leading to metallic behavior. Hybridization between Na($s$) and C($p_z$) orbitals dominates near the Fermi level. For clarity, the PDOS curves corresponding to the Na orbitals were magnified to enhance their visibility in the plot.}
    \label{fig:band_dos_na}
\end{figure*}

The effect of sodium decoration on the electronic structure of Na@P-C$_3$N is shown in  Figure~\ref{fig:band_dos_na}. The band structure (Figure~\ref{fig:band_dos_na}a) reveals the emergence of two cone-shaped points tilted at the Fermi level ($E_F$, represented by the red dashed line). Some modifications are noticeable in the band structure, such as the $E_F$, which has been shifted for the conduction band region, considering the band structure for the pristine monolayer. One can note that the band dispersion is very close to the one reported at VBM for the pristine configuration. This shift $E_F$ means that previously unoccupied states are now occupied.

The PDOS analysis (Figure~\ref{fig:band_dos_na}b) shows that the lower valence band of the Na-decorated P-C$_3$N retains the same orbital characteristics as the pristine structure. In the energy range between --2.0 eV and --1.5 eV, the density of states is mainly composed of C($p_z$) and N($p_z$) orbitals, while other atomic orbitals make negligible contributions. A distinct energy gap appears between --1.5 eV and --1.0 eV, corresponding to the band gap of the pristine monolayer.

Near the Fermi level (0 eV), Na($s$) states begin to contribute noticeably, while the amount of N($p_z$) states diminishes. This shift indicates that the C($p_z$) orbitals play a central role in mediating the interaction between Na adatoms and the monolayer, acting as primary acceptor states for the transferred charge. The emergence of Na-derived states at the Fermi level reflects a substantial charge transfer from Na atoms to the substrate, consistent with strong chemisorption and effective electron donation. 

\begin{figure}[!htb]
    \centering
    \includegraphics[width=0.3\linewidth]{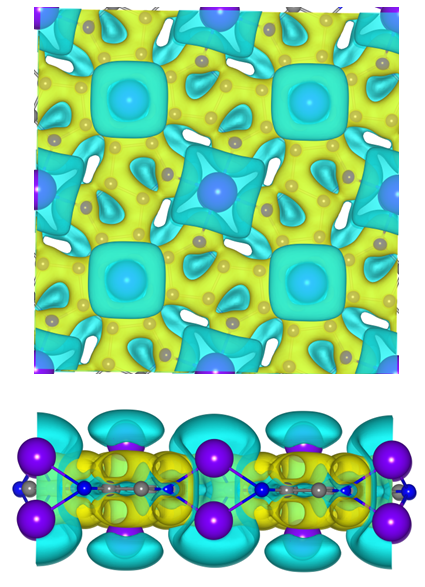}
    \caption{Charge density difference (CDD) for Na@P-C$_3$N. Yellow and cyan regions represent electron accumulation and depletion, respectively. Na atoms donate charge to the surrounding nitrogen framework.}
    \label{fig:cdd_na}
\end{figure}

 Figure~\ref{fig:cdd_na} presents the top and side views of the charge density difference (CDD) plots to gain insight into charge redistribution in the Na@P-C$_3$N system. Here, the yellow (cyan) indicates accumulation (depletion) charge regions. The zone of charge accumulation in the monolayer plane is observed, which denotes the charge transference from the Na adatoms to the P-C$_3$N monolayer. Significant regions of charge depletion are noticed around the Na adatoms. To corroborate this, Bader charge analysis was used, which revealed a charge accumulation of $-2.97~\svert e\svert $ on the P-C$_3$N monolayer, while the Na adatoms at the P1 site donated $-0.71~\svert e\svert $ and those at the P3 site $-0.78~\svert e\svert $. These charge transfers result in a strongly electrostatic interaction with Na adatoms and the P-C$_3$N. This redistribution enhances the electrostatic environment, allowing the induction of dipoles on H$_2$ and later physisorption, which is ideal for reversible hydrogen storage.

\begin{figure*}[!htb]
    \centering
    \includegraphics[width=0.7\linewidth]{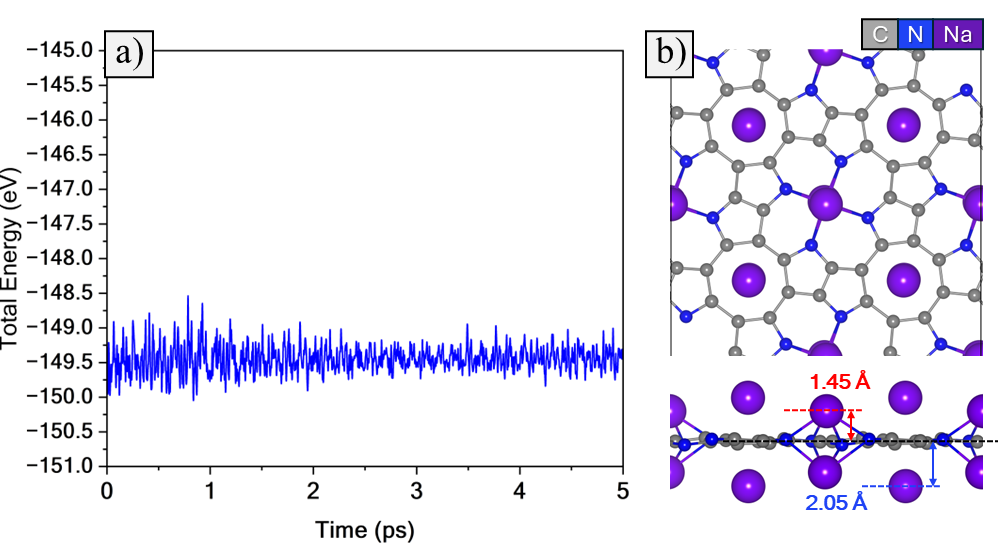}
    \caption{(a) AIMD total energy profile of Na@P-C$_3$N at 300~K over 5~ps. (b) The final structure shows that the Na atoms remain at the pore sites with minimal displacement. Distances of 1.45~\r{A} and 2.05~\r{A} are maintained relative to the surrounding framework.}
    \label{fig:aimd_na}
\end{figure*}

The thermal stability of the Na@P-C$_3$N complex was evaluated using AIMD simulations at 300~K for 5~ps, as shown in Figure~\ref{fig:aimd_na}. The total energy fluctuates within a narrow range without signs of structural degradation or Na diffusion. In addition, it is interesting to note that the distance between Na adatoms and the P-C$_3$N monolayer increased from 1.23~\AA~to 1.45~\AA~and 1.87~\AA~to 2.05~\AA~for P1 and P3 sites, respectively. These variations do not significantly affect the interaction between the Na adatoms and the monolayer, as can be evidenced by the fact that the Na atoms remain accommodated on their adsorption sites. The preservation of Na atoms at their pore-centered sites throughout the trajectory demonstrates that the system remains thermodynamically stable under near-ambient conditions.

\subsection{Hydrogen Storage Performance of Na@P-C$_3$N}

\begin{figure*}[!htb]
    \centering
    \includegraphics[width=0.4\linewidth]{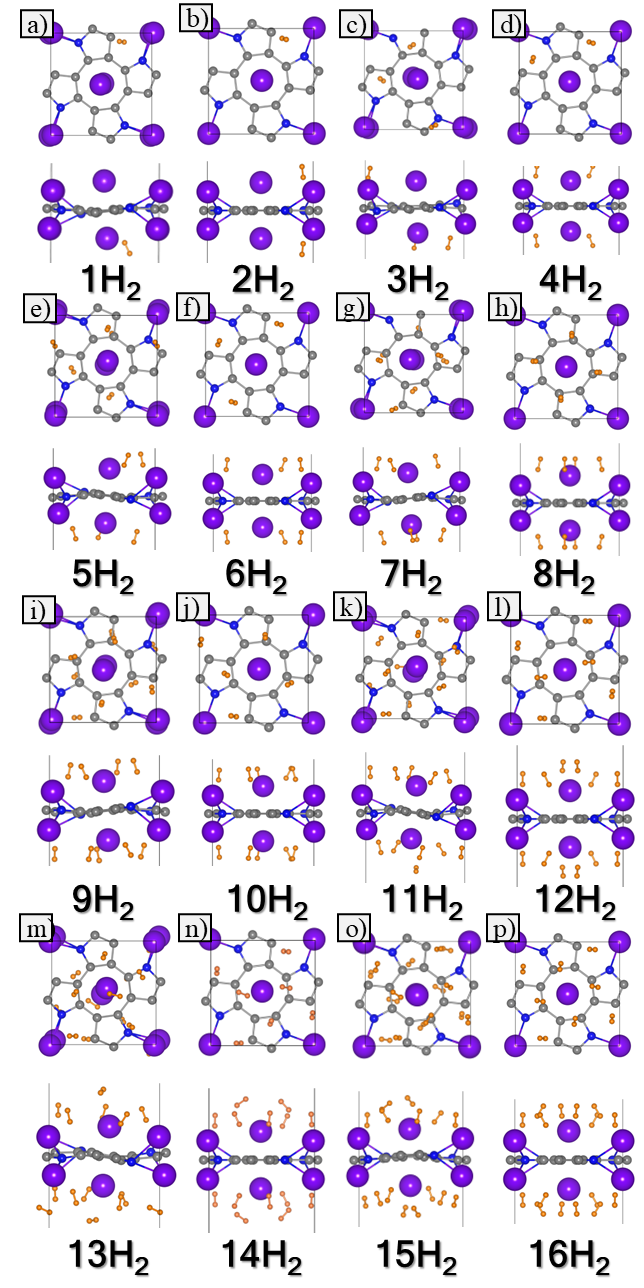}
    \caption{Optimized top and side views of Na@P-C$_3$N upon sequential adsorption of 1 to 16 H$_2$ molecules. Hydrogen molecules (orange) remain intact and symmetrically distributed around Na sites.}
    \label{fig:h2_structures}
\end{figure*}

The adsorption behavior of molecular hydrogen in the Na-decorated P-C$_3$N system was investigated by incrementally adding H$_2$ molecules to the optimized Na@P-C$_3$N structure.  Figure~\ref{fig:h2_structures} displays the relaxed configurations of 1 to 16 adsorbed H$_2$ molecules. The top and side views reveal that H$_2$ molecules preferentially adsorb around the N-bonded Na adatoms, maintaining their molecular character without dissociation, as expected by physisorption interaction. The increasing number of H$_2$ units leads to a homogeneous distribution around the adsorption centers, without causing significant structural deformation in the underlying framework. 

Table~\ref{tab:h2_adsorption_extended} summarizes the key parameters governing hydrogen adsorption on Na@P-C$_3$N, including the adsorption energy (E$_{\mathrm{ads}}$), average H--H bond length (R$_{\mathrm{H-H}}$), desorption temperature (T$_{\mathrm{D}}$), and hydrogen storage capacity (HSC). The calculated adsorption energies range from --0.18 to --0.28~eV per H$_2$ molecule, falling within the ideal energy window for reversible physisorption. 

The H--H bond length remains nearly constant at approximately 0.76--0.77~\AA, indicating minimal bond elongation and preserving the molecular nature of adsorbed H$_2$, a hallmark of physisorption. This is further corroborated by the calculated desorption temperatures, which range from 231 to 357~K based on the van't Hoff equation. These values are compatible with practical operating conditions and suggest thermally accessible hydrogen release at or near room temperature.

Importantly, the hydrogen storage capacity reaches a maximum of 9.88~wt\% for 16 H$_2$ molecules per unit cell. This value surpasses the U.S. Department of Energy (DOE) 2025 target of 6.5~wt\% for on-board hydrogen storage systems, highlighting the potential of Na-decorated P-C$_3$N as a high-capacity hydrogen storage material.

Furthermore, no abrupt changes in the adsorption energy are observed as more H$_2$ molecules are adsorbed, suggesting uniform and sequential physisorption rather than site saturation or clustering effects. The slight reduction in E$_{\mathrm{ads}}$ in higher coverage is expected due to weak H$_2$--H$_2$ repulsion and reduced availability of optimal adsorption sites.

To further understand the stepwise adsorption behavior, the consecutive adsorption energy ($E_{\mathrm{con}}$) was also calculated and is listed in \tabref{tab:h2_adsorption_extended}. This parameter reflects the incremental energy cost or gain associated with the adsorption of each successive H$_2$ molecule. Ideally, a smoothly decreasing or nearly constant $E_{\mathrm{con}}$ indicates a uniform and favorable adsorption process.

In the present case, although the majority of $E_{\mathrm{con}}$ values are negative --- indicating thermodynamically favorable stepwise adsorption --- some irregularities appear at specific coverage levels (e.g.,~for the 6th, 12th, and 14th H$_2$ molecules), where slightly positive values are observed. These deviations may be attributed to multiple factors.

First, due to the complexity of the system and the numerous possible adsorption geometries, the geometry chosen for each additional H$_2$ may not necessarily correspond to the global minimum energy configuration. In other words, the optimization may have converged to a local minimum that is less favorable energetically, resulting in a non-monotonic trend in $E_{\mathrm{con}}$. This is a common feature in high-coverage physisorption systems, where the spatial distribution of adsorbed molecules can lead to subtle repulsive interactions or steric hindrance.

\begin{table*}[h!]
\centering
\caption{Adsorption energy ($E_\mathrm{ads}$), average H--H bond length ($R_\mathrm{H-H}$), desorption temperature ($T_D$), consecutive adsorption energy ($E_\mathrm{con}$), and hydrogen storage capacity (HSC) for Na@P-C$_3$N with different numbers of H$_2$ molecules.}
\label{tab:h2_adsorption_extended}
\begin{tabular}{lccccc}
\hline
\textbf{System} & $E_\mathrm{ads}$ (eV) & $R_\mathrm{H-H}$ (\AA) & $T_D$ (K) & $E_\mathrm{con}$ (eV) & HSC (wt\%) \\
\hline
Na@P-C$_3$N--1H$_2$  & --0.27 & 0.77 & 342.51 & -- &  0.68 \\
Na@P-C$_3$N--2H$_2$  & --0.22 & 0.76 & 285.15 & --0.18 & 1.35 \\
Na@P-C$_3$N--3H$_2$  & --0.28 & 0.77 & 356.82 & --0.39 & 2.01 \\
Na@P-C$_3$N--4H$_2$  & --0.23 & 0.76 & 292.05 & --0.08 & 2.67 \\
Na@P-C$_3$N--5H$_2$  & --0.27 & 0.76 & 351.56 & --0.46 & 3.31 \\
Na@P-C$_3$N--6H$_2$  & --0.23 & 0.76 & 289.83 & 0.01 & 3.95 \\
Na@P-C$_3$N--7H$_2$  & --0.27 & 0.76 & 346.35 & --0.54 & 4.58 \\
Na@P-C$_3$N--8H$_2$  & --0.26 & 0.76 & 335.89 & --0.21 & 5.19 \\
Na@P-C$_3$N--9H$_2$  & --0.25 & 0.76 & 320.01 & --0.15 & 5.80 \\
Na@P-C$_3$N--10H$_2$ & --0.24 & 0.76 & 304.59 & --0.13 & 6.41 \\
Na@P-C$_3$N--11H$_2$ & --0.23 & 0.76 & 295.65 & --0.16 & 7.00 \\
Na@P-C$_3$N--12H$_2$ & --0.20 & 0.76 & 260.72 & 0.10 & 7.59 \\
Na@P-C$_3$N--13H$_2$ & --0.22 & 0.76 & 276.58 & --0.37 & 8.18 \\
Na@P-C$_3$N--14H$_2$ & --0.19 & 0.76 & 248.04 & 0.10 & 8.75 \\
Na@P-C$_3$N--15H$_2$ & --0.19 & 0.76 & 244.38 & --0.15 & 9.32 \\
Na@P-C$_3$N--16H$_2$ & --0.18 & 0.76 & 231.23 & --0.03 & 9.88 \\
\hline
\end{tabular}
\end{table*}

Attempts to adsorb a 17th H$_2$ molecule resulted in either complete desorption after relaxation or negligible adsorption energy (close to 0 eV), suggesting that no further interaction is thermodynamically favored. This behavior is attributed to steric hindrance and repulsive interactions between adjacent H$_2$ molecules, as well as the saturation of available electronic states around the Na adsorption centers. Therefore, 16 H$_2$ molecules per unit cell represent the practical saturation limit for this system.

\begin{figure*}[!htb]
    \centering
    \includegraphics[width=\linewidth]{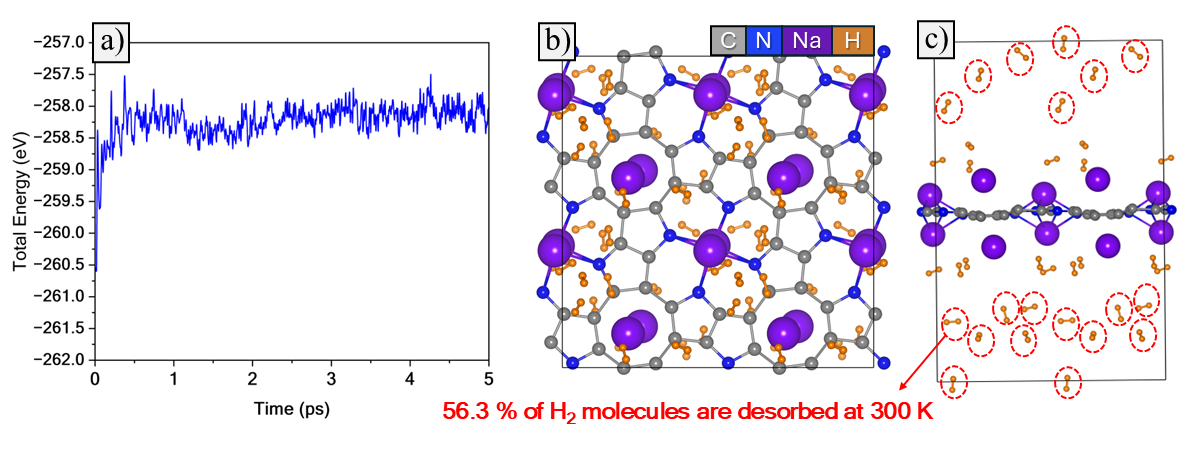}
    \caption{(a) Total energy profile during AIMD simulation of H$_2$-saturated Na@P-C$_3$N at 300~K. (b, c) Final configuration after 5~ps, highlighting partial desorption of H$_2$ (red dashed circles), with 56.3\% of molecules retained on the surface.}
    \label{fig:aimd_h2}
\end{figure*}

To assess the thermal stability and desorption behavior of the H$_2$-saturated system, AIMD simulations were performed at 300~K for 5~ps. Figure~\ref{fig:aimd_h2}a shows the evolution of total energy over time, which remains stable after an initial equilibration period. The final configuration (Figure~\ref{fig:aimd_h2}b, c) reveals partial desorption of hydrogen molecules, consistent with a thermodynamically reversible adsorption process. A quantitative analysis shows that approximately 56.3\% of the adsorbed H$_2$ molecules are released from the Na@P-C$_3$N substrate after 5~ps.

Although the AIMD simulations were limited to 5 ps due to computational constraints, the results already confirm structural and thermal stability at room temperature. No evidence of Na diffusion or H$_2$ detachment was observed during the simulations. However, we acknowledge that longer timescale simulations or higher-temperature studies may provide deeper insight into dynamic processes such as metal adatom mobility or thermally activated desorption.

\begin{figure*}[!htb]
    \centering
    \includegraphics[width=0.7\linewidth]{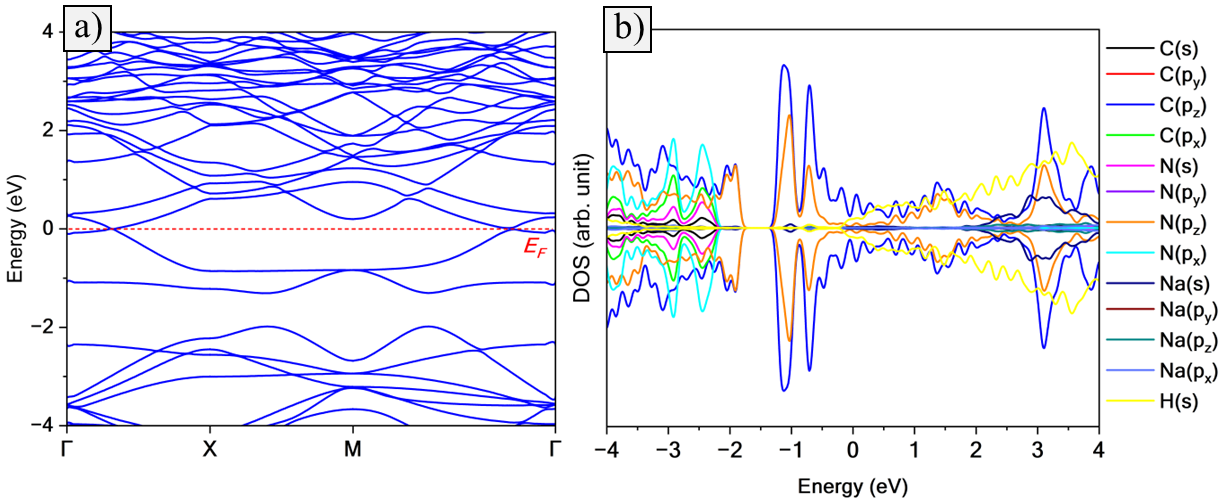}
    \caption{(a) Electronic band structure and (b) PDOS for H$_2$-saturated Na@P-C$_3$N. H($s$) orbitals contribute near the Fermi level, reflecting non-dissociative interactions.}
    \label{fig:band_dos_h2}
\end{figure*}

The electronic properties of the H$_2$-adsorbed Na@P-C$_3$N system, shown in  Figure~\ref{fig:band_dos_h2}, confirm the robustness of its semimetallic character even under full hydrogenation. The band structure (Figure~\ref{fig:band_dos_h2}a) reveals the preservation of the characteristic cone-shaped crossings at the Fermi level ($E_F$), indicating that the electronic symmetry and charge transport pathways remain essentially unchanged. The projected density of states (PDOS) in Figure~\ref{fig:band_dos_h2}b shows minimal changes compared to the pristine Na@P-C$_3$N system, with new contributions from H($s$) orbitals emerging near the Fermi level and extending into the conduction band region. These contributions are weak, reinforcing the physisorption nature of H$_2$ molecules and suggesting minimal hybridization between hydrogen and the substrate.

\begin{figure}[!htb]
    \centering
    \includegraphics[width=0.3\linewidth]{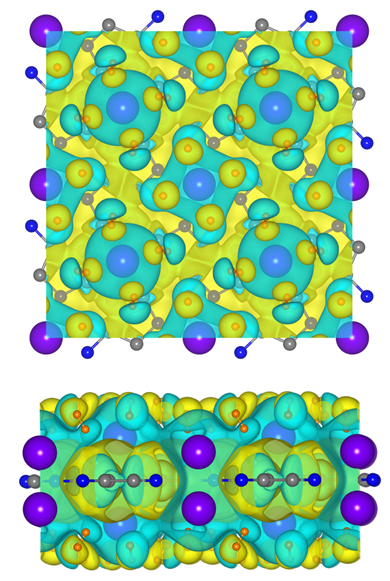}
    \caption{CDD plots of the fully hydrogenated Na@P-C$_3$N system. Cyan and yellow isosurfaces represent charge depletion and accumulation, respectively.}
    \label{fig:cdd_h2}
\end{figure}

The charge density difference (CDD) analysis provides additional information on the nature of the H$_2$ interactions represented by Figure~\ref{fig:cdd_h2}. It can be seen that the introduction of H$_2$ reinforces the charge transference previously reported between the Na adatoms and the P-C$_3$N, which accumulates charge regions between Na and the monolayer. To corroborate this, the Bader charge analysis shows that the N-coordinated Na atoms now have a charge of $+0.81\svert e\svert $, while those in the octagonal pore have a charge of $+0.78\svert e\svert $. The charge of Na atoms at the P1 site is higher than that reported for the Na@P-C$_3$N isolated complex of $+0.71\svert e\svert $. This shows that $N$ atoms in pyrrolopyrrole motifs are related to the charge accumulation on the P-C$_3$N after hydrogenation. This indicates that the N-coordinated sites are primarily responsible for stabilizing the adsorbed hydrogen.
Furthermore, simultaneous zones of charge depletion and accumulation are observed along the bond axes for H$_2$ molecules, which denote the induced dipole in these molecules. However, some charge accumulation is verified between Na adatoms and H$_2$ molecules. This is consistent with Bader charge analysis, which reveals charge transfer of $-0.02\svert e\svert $ per molecule from the substrate to H$_2$ molecules. 

This minimal charge redistribution (\textminus{}0.02 e per H$_2$ molecule from Na atoms to the adsorbed hydrogen) indicates that the H$_2$ molecules remain in a nearly neutral state, preserving their molecular character and preventing bond activation. Such weak physisorption interactions are desirable for reversible hydrogen storage, as they enable facile desorption without requiring high thermal input or chemical recombination.

\begin{figure}[!htb]
    \centering
    \includegraphics[width=0.5\linewidth]{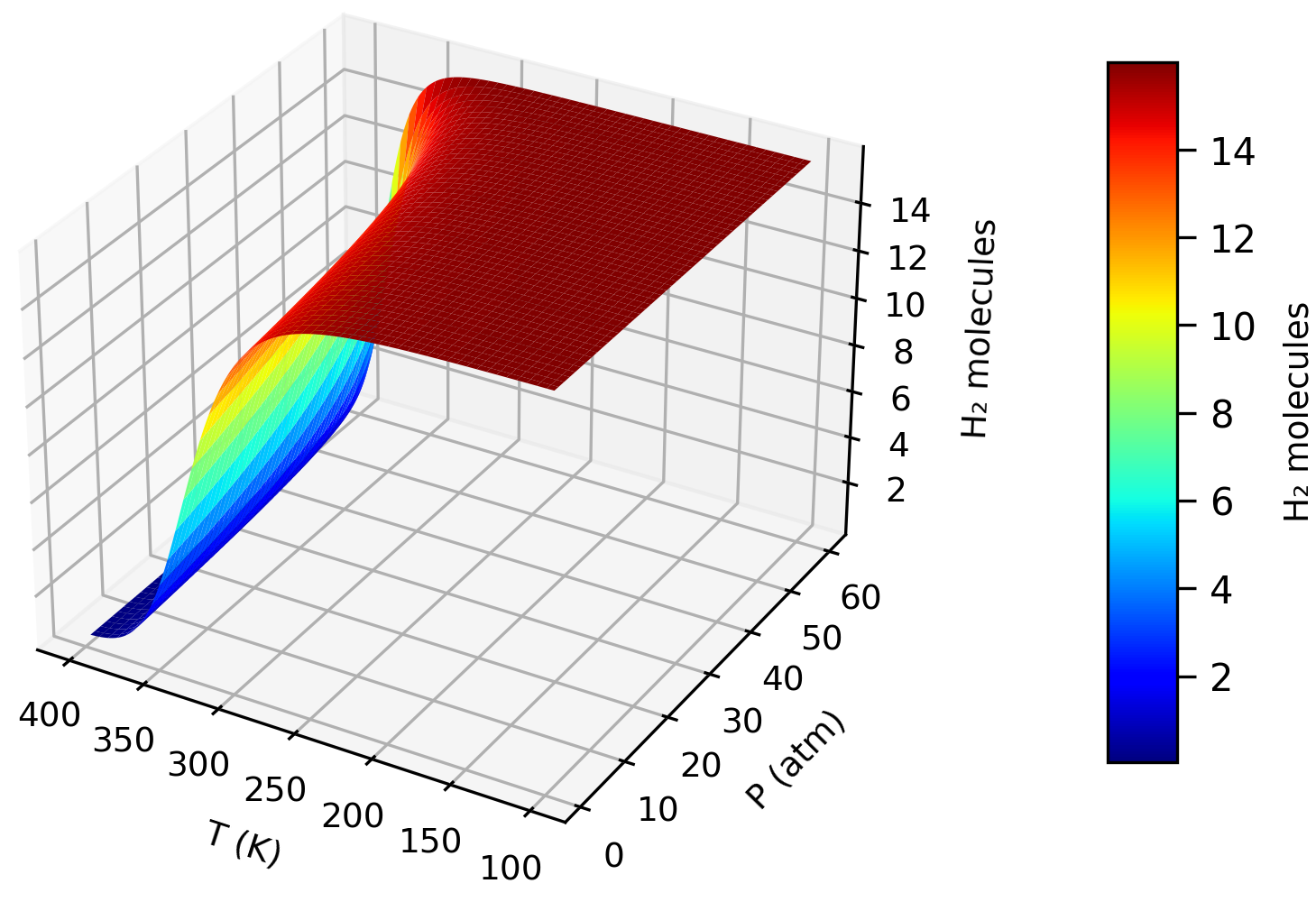}
    \caption{Thermodynamic response surface showing the number of adsorbed H$_2$ molecules on Na@P-C$_3$N as a function of temperature and pressure. The red plateau indicates full occupation. The blue region corresponds to desorption.}
    \label{fig:thermo_surface}
\end{figure}

To evaluate the practical usability of Na@P-C$_3$N for hydrogen storage, we calculated the occupancy of H$_2$ molecules as a function of temperature and pressure using a grand canonical ensemble approach. The resulting adsorption isotherm surface is shown in  Figure~\ref{fig:thermo_surface}. The study considered hydrogen uptake at 25~${\degree} {\rm C}$, 30~atm, and release at 100~${\degree} {\rm C}$, 3~atm. The storage capacity was determined by the difference in H$_2$ molecules under these conditions. Results indicate 15.62 H$_2$ molecules are adsorbed at low temperature/high pressure, with 0.13 H$_2$ remaining under desorption. This corresponds to a 9.59~wt\% gravimetric hydrogen storage, surpassing the U.S. Department of Energy target of 6.5~wt\%.

To benchmark the performance of Na@P-C$_3$N, we compare it with similar metal-decorated systems recently reported in the literature. Table~\ref{table:adsorption1} shows that Na@P-C$_3$N achieves a competitive combination of high storage capacity (9.88~wt\%) and moderate adsorption energy (0.18~eV), with a desorption temperature well suited for a near-ambient operation.

\begin{table*}[!ht]
\centering
\caption{Total number of adsorbed H$_2$ molecules ($n$), Absolute adsorption energy per H$_2$ (|$\text{E}_\text{ads}$|), Hydrogen Storage Capacity (HSC), and Desorption temperature (T$_\text{des}$) associated with configurations exhibiting complete H$_2$ coverage configurations in recently documented systems.} 
\label{table:adsorption1}
\begin{tabular*}{\linewidth}{@{\extracolsep{\fill}} lcccc @{}}
\toprule
\textbf{System}                      & \textbf{n} & \textbf{|$\text{E}_\text{ads}$| (eV)} & \textbf{HSC (wt\%)} & \textbf{T$_\text{des}$ (K)} \\
\midrule
\textbf{Na@P-C$_3$N} (this work) & 16 & 0.18  & 9.88  & 231  \\
\textbf{Li@$\alpha$-C$_3$N$_2$} \cite{CHEN2024510} & 12 & 0.215  & 5.7  & 277 \\
\textbf{Na@B$_7$N$_5$} \cite{LIU2025105802} & 32  & 0.20  & 7.70  & 257  \\ 
\textbf{Na@Irida-graphene} \cite{duan2024reversible} & 32 & 0.14 & 7.82 & 195 \\
\textbf{Na@IGP-SiC} \cite{MARTINS202498} & 48 & 0.10 & 6.78 & 148 \\
\textbf{K@BP-Biphenylene} \cite{djebablia2024metal} & 32 & 0.14 & 8.27 & - \\
\textbf{NLi$_4$@Phosphorene} \cite{boubkri2024computational} & 30 & 0.11 & 6.8 & 82 \\
\textbf{Li@POG-B$_4$$\alpha$-C$_3$N$_2$$_3$} \cite{chen2025penta} & 10 & 0.19 & 8.35 & 245 \\
\textbf{K@PHE-graphene} \cite{LARANJEIRA2025139} & 32 & 0.33 & 7.47 & 423 \\
\bottomrule
\end{tabular*}
\end{table*}

\section{Conclusions}

In this study, we proposed and evaluated Na-decorated P-C$_3$N as a promising 2D material for reversible hydrogen storage applications. DFT calculations confirmed that sodium atoms strongly adsorb in the center of intrinsic pores of P-C$_3$N, with high binding energy ($-4.48~{\rm eV}$) and no lattice deformation. Upon decoration, the system transitions from semiconducting to metallic, maintaining excellent thermal and structural stability at 300~K.

Sequential adsorption of H$_2$ molecules revealed physisorption interactions with adsorption energies ranging from $-0.18$ to $-0.28~{\rm eV}$ and minimal H--H bond elongation, indicating non-dissociative, electrostatically driven binding. A maximum of 16 hydrogen molecules could be adsorbed per unit cell, yielding a high gravimetric capacity of 9.88~wt\% and practical storage of 9.59~wt\% under realistic conditions. Charge analysis showed weak electron transfer to H$_2$, primarily from sodium atoms coordinated to nitrogen, supporting the observed reversibility.

Compared to other metal-decorated 2D materials, Na@P-C$_3$N offers a balanced combination of storage capacity, moderate adsorption strength, and ambient-operable desorption temperatures. These results establish Na@P-C$_3$N as a viable platform for lightweight, regenerable hydrogen storage and motivate further experimental validation.

\section*{Data access statement}
Data supporting the results can be accessed by contacting the corresponding author.

\section*{Conflicts of interest}
The authors declare no conflict of interest.

\section*{Acknowledgements}
This work was supported by the Brazilian funding agencies Fundação de Amparo à Pesquisa do Estado de São Paulo - FAPESP (grant no. 2022/03959-6, 2022/00349- 2, 2022/14576-0, 2020/01144-0, 2024/05087-1, and 2022/16509-9), and National Council for Scientific and Technological Development - CNPq (grant no. 307213/2021–8). L.A.R.J. acknowledges the financial support from FAP-DF grants 00193.00001808/2022-71 and $00193-00001857/2023-95$, FAPDF-PRONEM grant 00193.00001247/2021-20, PDPG-FAPDF-CAPES Centro-Oeste 00193-00000867/2024-94, and CNPq grants $350176/2022-1$ and $167745/2023-9$. K.A.L.L. acknowledge the Center for Computational Engineering \& Sciences (CCES) at Unicamp for financial support through the FAPESP/CEPID Grant 2013/08293-7. X.C. was funded by the Research Program of Chongqing Municipal Education Commission (No. KJQN202201327 and No. KJQN202301339), and the Natural Science Foundation of Chongqing, China (CSTB2022NSCQ-MSX0621).

\section*{Declaration of generative AI and AI-assisted technologies in the writing process}
During the preparation of this work, the authors used Writefull to improve readability and language. After using this tool/service, the authors reviewed and edited the content as needed and assume full responsibility for the content of the publication.

\printcredits

\bibliography{cas-refs}

\end{document}